\documentclass[letterpaper,aps,prc,superscriptaddress,showpacs,nofootinbib,floatfix,twocolumn]{revtex4}

\usepackage{graphicx}	
\usepackage{multirow}
\usepackage{amssymb}
\usepackage{amsmath}
\usepackage{xspace}	
\usepackage{dcolumn}
\usepackage{bm}

\newcommand{\pt}{\mbox{$p_T$}\xspace}

\newcommand{\Np}{\mbox{$N_{\rm part}$}\xspace}

\newcommand \sqsn{\mbox{$\sqrt{s_{_{NN}}}$}\xspace}

\begin{document}

\title{Nuclear modification factors of $\phi$ meson 
       in $d$+Au, Cu+Cu and Au+Au collisions at \sqsn= 200 GeV}

\newcommand{\abilene}{Abilene Christian University, Abilene, Texas 79699, USA}
\newcommand{\banaras}{Department of Physics, Banaras Hindu University, Varanasi 221005, India}
\newcommand{\barc}{Bhabha Atomic Research Centre, Bombay 400 085, India}
\newcommand{\bnlcoll}{Collider-Accelerator Department, Brookhaven National Laboratory, Upton, New York 11973-5000, USA}
\newcommand{\bnlphys}{Physics Department, Brookhaven National Laboratory, Upton, New York 11973-5000, USA}
\newcommand{\caucr}{University of California - Riverside, Riverside, California 92521, USA}
\newcommand{\charlesczech}{Charles University, Ovocn\'{y} trh 5, Praha 1, 116 36, Prague, Czech Republic}
\newcommand{\chonbuk}{Chonbuk National University, Jeonju, 561-756, Korea}
\newcommand{\ciae}{China Institute of Atomic Energy (CIAE), Beijing, People's Republic of China}
\newcommand{\cns}{Center for Nuclear Study, Graduate School of Science, University of Tokyo, 7-3-1 Hongo, Bunkyo, Tokyo 113-0033, Japan}
\newcommand{\colorado}{University of Colorado, Boulder, Colorado 80309, USA}
\newcommand{\columbia}{Columbia University, New York, New York 10027 and Nevis Laboratories, Irvington, NY 10533, USA}
\newcommand{\czechtech}{Czech Technical University, Zikova 4, 166 36 Prague 6, Czech Republic}
\newcommand{\dapnia}{Dapnia, CEA Saclay, F-91191, Gif-sur-Yvette, France}
\newcommand{\debrecen}{Debrecen University, H-4010 Debrecen, Egyetem t{\'e}r 1, Hungary}
\newcommand{\elte}{ELTE, E{\"o}tv{\"o}s Lor{\'a}nd University, H - 1117 Budapest, P{\'a}zm{\'a}ny P. s. 1/A, Hungary}
\newcommand{\ewha}{Ewha Womans University, Seoul 120-750, Korea}
\newcommand{\fit}{Florida Institute of Technology, Melbourne, Florida 32901, USA}
\newcommand{\fsu}{Florida State University, Tallahassee, Florida 32306, USA}
\newcommand{\gsu}{Georgia State University, Atlanta, Georgia 30303, USA}
\newcommand{\hiroshima}{Hiroshima University, Kagamiyama, Higashi-Hiroshima 739-8526, Japan}
\newcommand{\ihepprot}{IHEP Protvino, State Research Center of Russian Federation, Institute for High Energy Physics, Protvino, 142281, Russia}
\newcommand{\illuiuc}{University of Illinois at Urbana-Champaign, Urbana, Illinois 61801, USA}
\newcommand{\instpasczech}{Institute of Physics, Academy of Sciences of the Czech Republic, Na Slovance 2, 182 21 Prague 8, Czech Republic}
\newcommand{\isu}{Iowa State University, Ames, Iowa 50011, USA}
\newcommand{\jinrdubna}{Joint Institute for Nuclear Research, 141980 Dubna, Moscow Region, Russia}
\newcommand{\jyvaskyla}{Helsinki Institute of Physics and University of Jyv{\"a}skyl{\"a}, P.O.Box 35, FI-40014 Jyv{\"a}skyl{\"a}, Finland}
\newcommand{\kaeri}{KAERI, Cyclotron Application Laboratory, Seoul, Korea}
\newcommand{\kek}{KEK, High Energy Accelerator Research Organization, Tsukuba, Ibaraki 305-0801, Japan}
\newcommand{\kfki}{KFKI Research Institute for Particle and Nuclear Physics of the Hungarian Academy of Sciences (MTA KFKI RMKI), H-1525 Budapest 114, POBox 49, Budapest, Hungary}
\newcommand{\korea}{Korea University, Seoul, 136-701, Korea}
\newcommand{\kurchatov}{Russian Research Center ``Kurchatov Institute", Moscow, Russia}
\newcommand{\kyoto}{Kyoto University, Kyoto 606-8502, Japan}
\newcommand{\labllr}{Laboratoire Leprince-Ringuet, Ecole Polytechnique, CNRS-IN2P3, Route de Saclay, F-91128, Palaiseau, France}
\newcommand{\lawllnl}{Lawrence Livermore National Laboratory, Livermore, California 94550, USA}
\newcommand{\losalamos}{Los Alamos National Laboratory, Los Alamos, New Mexico 87545, USA}
\newcommand{\lpc}{LPC, Universit{\'e} Blaise Pascal, CNRS-IN2P3, Clermont-Fd, 63177 Aubiere Cedex, France}
\newcommand{\lund}{Department of Physics, Lund University, Box 118, SE-221 00 Lund, Sweden}
\newcommand{\maryland}{University of Maryland, College Park, Maryland 20742, USA}
\newcommand{\mass}{Department of Physics, University of Massachusetts, Amherst, MA 01003-9337, USA }
\newcommand{\muenster}{Institut fur Kernphysik, University of Muenster, D-48149 Muenster, Germany}
\newcommand{\muhlenberg}{Muhlenberg College, Allentown, Pennsylvania 18104-5586, USA}
\newcommand{\myongji}{Myongji University, Yongin, Kyonggido 449-728, Korea}
\newcommand{\nagasaki}{Nagasaki Institute of Applied Science, Nagasaki-shi, Nagasaki 851-0193, Japan}
\newcommand{\newmex}{University of New Mexico, Albuquerque, New Mexico 87131, USA }
\newcommand{\nmsu}{New Mexico State University, Las Cruces, New Mexico 88003, USA}
\newcommand{\ornl}{Oak Ridge National Laboratory, Oak Ridge, Tennessee 37831, USA}
\newcommand{\orsay}{IPN-Orsay, Universite Paris Sud, CNRS-IN2P3, BP1, F-91406, Orsay, France}
\newcommand{\peking}{Peking University, Beijing, People's Republic of China}
\newcommand{\pnpi}{PNPI, Petersburg Nuclear Physics Institute, Gatchina, Leningrad region, 188300, Russia}
\newcommand{\riken}{RIKEN Nishina Center for Accelerator-Based Science, Wako, Saitama 351-0198, JAPAN}
\newcommand{\rikjrbrc}{RIKEN BNL Research Center, Brookhaven National Laboratory, Upton, New York 11973-5000, USA}
\newcommand{\rikkyo}{Physics Department, Rikkyo University, 3-34-1 Nishi-Ikebukuro, Toshima, Tokyo 171-8501, Japan}
\newcommand{\saispbstu}{Saint Petersburg State Polytechnic University, St. Petersburg, Russia}
\newcommand{\saopaulo}{Universidade de S{\~a}o Paulo, Instituto de F\'{\i}sica, Caixa Postal 66318, S{\~a}o Paulo CEP05315-970, Brazil}
\newcommand{\seoulnat}{System Electronics Laboratory, Seoul National University, Seoul, Korea}
\newcommand{\stonybrkc}{Chemistry Department, Stony Brook University, Stony Brook, SUNY, New York 11794-3400, USA}
\newcommand{\stonycrkp}{Department of Physics and Astronomy, Stony Brook University, SUNY, Stony Brook, New York 11794, USA}
\newcommand{\subatech}{SUBATECH (Ecole des Mines de Nantes, CNRS-IN2P3, Universit{\'e} de Nantes) BP 20722 - 44307, Nantes, France}
\newcommand{\tenn}{University of Tennessee, Knoxville, Tennessee 37996, USA}
\newcommand{\titech}{Department of Physics, Tokyo Institute of Technology, Oh-okayama, Meguro, Tokyo 152-8551, Japan}
\newcommand{\tsukuba}{Institute of Physics, University of Tsukuba, Tsukuba, Ibaraki 305, Japan}
\newcommand{\vandy}{Vanderbilt University, Nashville, Tennessee 37235, USA}
\newcommand{\waseda}{Waseda University, Advanced Research Institute for Science and Engineering, 17 Kikui-cho, Shinjuku-ku, Tokyo 162-0044, Japan}
\newcommand{\weizmann}{Weizmann Institute, Rehovot 76100, Israel}
\newcommand{\yonsei}{Yonsei University, IPAP, Seoul 120-749, Korea}
\affiliation{\abilene}
\affiliation{\banaras}
\affiliation{\barc}
\affiliation{\bnlcoll}
\affiliation{\bnlphys}
\affiliation{\caucr}
\affiliation{\charlesczech}
\affiliation{\chonbuk}
\affiliation{\ciae}
\affiliation{\cns}
\affiliation{\colorado}
\affiliation{\columbia}
\affiliation{\czechtech}
\affiliation{\dapnia}
\affiliation{\debrecen}
\affiliation{\elte}
\affiliation{\ewha}
\affiliation{\fit}
\affiliation{\fsu}
\affiliation{\gsu}
\affiliation{\hiroshima}
\affiliation{\ihepprot}
\affiliation{\illuiuc}
\affiliation{\instpasczech}
\affiliation{\isu}
\affiliation{\jinrdubna}
\affiliation{\jyvaskyla}
\affiliation{\kaeri}
\affiliation{\kek}
\affiliation{\kfki}
\affiliation{\korea}
\affiliation{\kurchatov}
\affiliation{\kyoto}
\affiliation{\labllr}
\affiliation{\lawllnl}
\affiliation{\losalamos}
\affiliation{\lpc}
\affiliation{\lund}
\affiliation{\maryland}
\affiliation{\mass}
\affiliation{\muenster}
\affiliation{\muhlenberg}
\affiliation{\myongji}
\affiliation{\nagasaki}
\affiliation{\newmex}
\affiliation{\nmsu}
\affiliation{\ornl}
\affiliation{\orsay}
\affiliation{\peking}
\affiliation{\pnpi}
\affiliation{\riken}
\affiliation{\rikjrbrc}
\affiliation{\rikkyo}
\affiliation{\saispbstu}
\affiliation{\saopaulo}
\affiliation{\seoulnat}
\affiliation{\stonybrkc}
\affiliation{\stonycrkp}
\affiliation{\subatech}
\affiliation{\tenn}
\affiliation{\titech}
\affiliation{\tsukuba}
\affiliation{\vandy}
\affiliation{\waseda}
\affiliation{\weizmann}
\affiliation{\yonsei}
\author{A.~Adare} \affiliation{\colorado}
\author{S.~Afanasiev} \affiliation{\jinrdubna}
\author{C.~Aidala} \affiliation{\columbia} \affiliation{\mass}
\author{N.N.~Ajitanand} \affiliation{\stonybrkc}
\author{Y.~Akiba} \affiliation{\riken} \affiliation{\rikjrbrc}
\author{H.~Al-Bataineh} \affiliation{\nmsu}
\author{J.~Alexander} \affiliation{\stonybrkc}
\author{A.~Al-Jamel} \affiliation{\nmsu}
\author{A.~Angerami} \affiliation{\columbia}
\author{K.~Aoki} \affiliation{\kyoto} \affiliation{\riken}
% ...
\collaboration{PHENIX Collaboration} \noaffiliation

\begin{abstract}


The PHENIX experiment at the Relativistic Heavy Ion Collider (RHIC) has 
performed systematic measurements of $\phi$ meson production in the 
K$^+$K$^-$ decay channel at midrapidity in $p+p$, $d$+Au, Cu+Cu and Au+Au 
collisions at \sqsn=200~GeV.  Results are presented on the $\phi$ invariant 
yield and the nuclear modification factor R$_{\rm AA}$ for Au+Au and Cu+Cu, 
and R$_{\rm dA}$ for $d$+Au collisions, studied as a function of transverse 
momentum ($1<p_T<7$~GeV/$c$) and centrality.  In central and 
mid-central Au+Au collisions, the R$_{\rm AA}$ of $\phi$ exhibits a 
suppression relative to expectations from binary scaled $p$+$p$ results.  
The amount of suppression is smaller than that of the $\pi^0$ and the 
$\eta$ in the intermediate $p_T$ range (2--5~GeV/$c$), whereas at 
higher $p_T$ the $\phi$, $\pi^0$ and $\eta$ show similar suppression.  
The baryon (protons and anti-protons) excess observed in central Au+Au 
collisions at intermediate $p_T$ is not observed for the $\phi$ meson 
despite the similar mass of the proton and the $\phi$.  This suggests that 
the excess is linked to the number of constituent quarks rather than the 
hadron mass.  The difference gradually disappears with decreasing 
centrality and for peripheral collisions the R$_{\rm AA}$ values for both 
particles are consistent with binary scaling.  Cu+Cu collisions show the 
same yield and suppression as Au+Au collisions for the same number of \Np.  
The R$_{\rm dA}$ of $\phi$ shows no evidence for cold nuclear effects 
within uncertainties.

\end{abstract}
 

\maketitle

\section{Introduction}

Measurements of hadron spectra from $p+p$ and nucleus-nucleus collisions at 
RHIC provide a means to study the mechanisms of particle production and the 
properties of the medium formed in relativistic heavy ion collisions.  At 
low transverse momentum, $p_T<2$~GeV/$c$, where the bulk of particles 
are produced, hadron production is governed by soft processes characterized 
by low momentum transfer.  The particle yields and the evolution of the 
interacting system are successfully described within the framework of 
thermal and hydrodynamical 
models~\cite{BraunMunzinger:2003zd,PHENIX_white,Kolb:2003dz,Jamie,Huovinen}.

At high transverse momentum, $p_T>5$ GeV/$c$, hard scattering 
processes become the dominant contribution.  Due to the large momentum 
transfer involved, the parton-parton scattering cross sections are amenable 
to perturbative QCD (pQCD) description and hadron production can be 
calculated using initial state parton distribution functions and final 
state fragmentation functions.  Modifications to the hadron yields are 
expected in nucleus-nucleus collisions due to the interaction of the 
scattered parton with the hot and dense medium 
formed~\cite{jetquenching,Baier1997,Gyulassy1994}.
In the absence of interaction with the medium the hard scatterings and the 
resulting hadron yields should scale with the number of binary 
nucleon-nucleon collisions ($N_{\rm coll}$), whereas in the medium the yields 
are suppressed (``jet-quenching'' \cite{jet_quench}) due to parton energy 
loss through gluon bremsstrahlung.  High-$p_T$ hadron suppression 
consistent with this scenario has been discovered in Au+Au collisions at 
RHIC~\cite{PPG003,PPG014,PPG084}.  The same suppression by a factor of 
$\sim5$ is observed for $\pi^{0}$ and $\eta$ production whereas direct 
photons which do not interact with the medium, follow the expected binary 
scaling~\cite{PPG042}.  Single electrons originating from the semi-leptonic 
decays of mesons containing heavy quarks (charm and bottom) exhibit a large 
suppression at high $p_T$, similar within the experimental 
uncertainties to that of $\pi^{0}$ and $\eta$, presenting a challenge for 
the bremsstrahlung explanation~\cite{heavy}.
 
At intermediate transverse momentum $2<p_T$~(GeV/$c$)$<5$, 
suppression of binary scaled production is observed for light $\pi^{0}$ and 
$\eta$ mesons but not for protons and anti-protons in mid-central and 
central Au+Au collisions~\cite{PPG015}.  The $p/\pi$ and $\bar{p}/\pi$ 
ratios increase with centrality and exceed the values measured in $p+p$ by 
a factor of 3--5 in the most central collisions.  A different suppression 
pattern between baryons and mesons is also observed for strange hadrons, 
$\Lambda$ and K$^{0}_{S}$~\cite{K_Lambda_STAR,Kstar_levy}.  These 
baryon/meson differences in suppression are inconsistent with the picture 
of hadron production through hard-scattering followed by partonic energy 
loss in medium and hadronization in vacuum according to the universal 
fragmentation functions.  This poses the question whether hard-scattering is 
the dominant source of baryon production at intermediate $p_T$.  
Studies of jet-like dihadron correlations in Au+Au collisions 
\cite{Adler2005c,PPG072} imply nearly equal importance of the jet 
fragmentation as a production mechanism for mesons and baryons, except for 
the most central collisions.  The interpretation of the baryon 
non-suppression results requires therefore another particle production 
mechanism in addition to jet fragmentation at intermediate $p_T$.

There have been attempts to describe the different behavior of baryons and 
mesons through the strong radial flow that boosts particles with larger 
mass to higher $p_T$~\cite{hydro2,hydro4} or through the 
recombination of soft and hard massive 
partons~\cite{recombination2,Hwa:2002tu,recombination1}, 
or through the interplay of the jet quenched hard component and 
phenomenological soft to moderate $p_T$ baryon junction 
component~\cite{Vitev2002} or through the QCD color transparency of 
higher-twist contributions to inclusive hadroproduction cross sections, 
where baryons are produced directly in short distance 
subprocess~\cite{brodsky}.  Although several models reproduce $p_T$ 
spectra, particle ratios and elliptic flow for different hadrons reasonably 
well, the relative contributions from the different processes are difficult 
to infer.

The $\phi$ meson is a very rich probe of the medium formed in heavy ion 
collisions, because it is sensitive to several aspects of the collision, 
including strangeness enhancement and chiral symmetry restoration, as 
well as energy loss and the nuclear modification 
factor~\cite{SomeReviewPaper,PPG016,STAR-phi2007,STAR-phi2009,PPG073}, 
which is the focus of this paper.  Due to its small inelastic cross section 
for interaction with non-strange hadrons~\cite{SomeReviewPaper,phi_cross2}, 
the $\phi$ meson is less affected by late hadronic rescattering and 
reflects better the initial evolution of the system.  Being a meson with a 
mass comparable to that of the proton, it is interesting to see how the 
$\phi$ meson fits within the meson/baryon pattern described previously; 
being a pure $s\bar{s}$ state, it puts additional constraints on the energy 
loss and recombination models.

This paper presents systematic PHENIX measurements of $\phi$ meson 
production via the K$^+$K$^-$ decay channel at \sqsn=200~GeV, including 
first PHENIX results in $p$+$p$, $d$+Au and Cu+Cu collisions and new 
results in Au+Au collisions.  The latter have much higher statistics and a 
finer centrality binning in comparison to the previously published PHENIX 
results \cite{PPG016}.  The results benefit from three different techniques 
involving different levels of kaon identification in the analyses.  These, 
combined with the high statistics of the analyzed data samples, allow for 
the extension of the $p_T$ range of the measurements up to $p_{\rm 
T}=7.0$~GeV/$c$ in all collision systems.  The higher $p_T$ reach and 
the higher precision of the data allow for sharper conclusions with respect 
to earlier results \cite{PPG016,STAR-phi2009}.  The Cu+Cu measurements are 
complementary to those on Au+Au and allow the study of nuclear effects with 
different nuclear overlap geometry for the same $N_{\rm part}$ and with smaller 
$N_{\rm part}$ uncertainties for $N_{\rm part}<100$.

The measurement of the $\phi$ meson production in $d$+Au collisions is 
important for understanding cold nuclear matter effects which are of 
interest by themselves and are also essential for the interpretation of 
heavy ion collisions.  As shown in~\cite{PPG030}, in the intermediate 
$p_T$ range, charged pions are practically not enhanced in comparison 
to the binary scaled $p+p$ yield, whereas protons and anti-protons exhibit 
some enhancement of $\sim$ 30\% in the most central collisions.  The 
mechanism of multiple soft re-scattering of partons in the initial state 
which is usually invoked as the origin of the Cronin effect does not 
explain this meson/baryon difference.  One possible explanation comes from 
recombination models~\cite{reco} in which baryons gain higher transverse 
momentum from recombination of three quarks in the final state in 
comparison to mesons consisting of only two quarks.  Measurement of the 
Cronin effect for the $\phi$ mesons can provide additional constraints for 
the models that try to explain these cold nuclear effects.

 

\section{Experimental set-up and data analysis}

We report on the measurements of $\phi$ mesons at midrapidity in the 
K$^+$K$^-$ decay channel in $p$+$p$, $d$+Au, Cu+Cu and Au+Au collisions at 
\sqsn=200~GeV using data collected by the PHENIX experiment during the 
2004, 2005 and 2008 physics runs.  A detailed description of the PHENIX 
detector can be found elsewhere~\cite{PHENIX}.  The measurements were done 
using the two PHENIX central arms each covering 90$^{\circ}$ in azimuth at 
midrapidity $(|\eta|<0.35)$.  The tracking of charged particles and the 
measurement of their momentum with typical resolution of $\delta p/p=0.7 
\oplus 1.1 \% p$~[GeV/$c$] are performed using the drift chambers and the 
first layer of the pad chambers (PC).  In order to reduce the background at 
high $p_T$, tracks are required to have a matching confirmation in 
the third layer of PC or the electromagnetic calorimeter (EMCal).  Kaons are 
identified using the time-of-flight detector (TOF), which covers 
approximately 1/3 of the acceptance in one of the central arms.  With a time 
resolution of $\sim$ 115~ps, the TOF allows for clear $\pi$/K separation in 
the range of transverse momentum from 0.3~GeV/$c$ to 2.2~GeV/$c$ using a 
2$\sigma$ $p_T$-dependent mass-squared selection cut as described 
in~\cite{PPG016}.

The beam-beam counters (BBC) and zero degree calorimeters (ZDC) are 
dedicated subsystems that determine the collision vertex along the beam 
axis ($z_{\rm vtx}$) and the event centrality, and also provide the minimum 
bias interaction trigger.  Events are categorized into centrality classes 
using two-dimensional cuts in the space of BBC charge versus ZDC 
energy~\cite{centralityAuAu} for Au+Au collisions or only by the amount of 
charge deposited in the BBC~\cite{centralitydAu,PPG084} for $d$+Au and 
Cu+Cu collisions.

In any particular event one cannot distinguish between kaons from $\phi$ 
decays and other kaons, so the $\phi$ meson yields are measured on a 
statistical basis.  In each event, all tracks of opposite charge which pass 
the selection criteria are paired to form the invariant mass distribution.  
This distribution contains both the signal (S) and an inherent 
combinatorial background (B).  To maximize the statistical significance and 
the \pt reach of the measurements we use three different track selection 
techniques: ``no PID'', in which all tracks are assigned the kaon mass, but 
no TOF information is used, and ``one kaon PID'' or ``two kaons PID'', in 
which one or both tracks are identified as kaons in the TOF.

\begin{table}[tbh]
 \caption{\label{nevt} 
Collision species, number of analyzed minimum bias events, 
accessible $p_T$ range and typical range of S/B ratio 
for the different $\phi\rightarrow$ K$^+$K$^-$ analyses.}
\begin{ruledtabular} \begin{tabular}{cccccc}
  Species &  & N [10$^{9}$] & $p_T$  [GeV/$c$] & S/B & Technique \\\hline 
\multirow{2}*{$p$+$p$} & \multirow{2}*{} & 1.50 &0.9$-$4.5& 1/9 - 1/2 & "one kaon PID" \\ 
& & 1.44 & 1.3$-$7.0& 1/76 - 1/3 & ``no PID'' \\ \\
  $d$+Au & & 1.69 & 1.1$-$7.0 & 1/245 - 1/12 & ``no PID'' \\ \\

\multirow{2}*{Cu+Cu} &\multirow{2}*{} & 0.77& 1.1$-$2.95 & 1/91 - 1/9 & ``one kaon PID'' \\
& & 0.78 &1.9$-$7.0 & 1/205 - 1/24 & ``no PID'' \\ \\

\multirow{2}*{Au+Au} &\multirow{2}*{} &
0.72 & 1.1$-$3.95 & 1/19 - 1/2 & ``two kaons PID'' \\ & &
0.82 &2.45$-$7.0 & 1/385 - 1/32 & ``no PID'' \\
\end{tabular}\end{ruledtabular}
 \end{table}

Table~\ref{nevt} lists for each collision system and for each analysis 
technique the number of analyzed minimum bias events in the vertex range 
$|z_{\rm vtx}|<30~{\rm cm}$, the accessible $p_T$ range and the range 
of the S/B ratio.

\begin{figure}[thb]
 \includegraphics[width=0.8\linewidth]{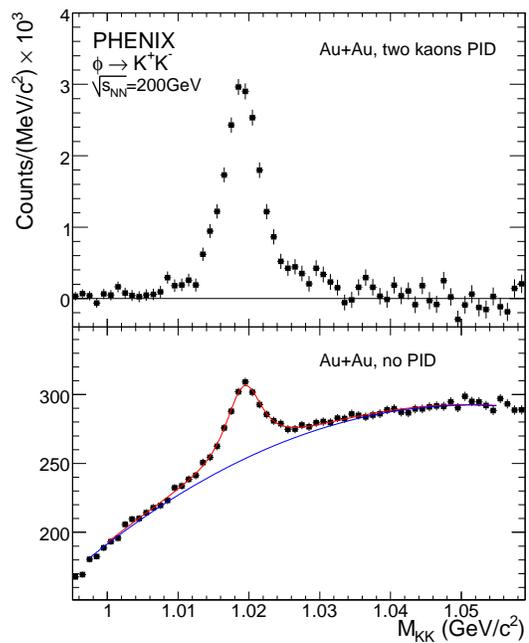} 
 \caption{\label{fig:Minv} (Color online) 
Invariant mass distributions obtained with the ``two kaons PID'' and ``no 
PID'' methods in Au+Au collisions after subtraction of the combinatorial 
background estimated using the event-mixing technique.  Plot on the top 
corresponds to integrated $p_T$ range whereas plot on the bottom is 
for the range $2<p_T$~(GeV/$c$)$<3$.  The ``no PID'' spectrum is fit 
to the sum of a Breit-Wigner function convolved with a Gaussian function to 
account for the $\phi$ signal, and a polynomial function to account for the 
residual background.
}
\end{figure}

The raw yields of the $\phi$ are obtained by integrating the invariant mass 
distributions in a window of $\pm$9~MeV/$c^{2}$ around the $\phi$ mass 
after subtracting the combinatorial background.  In the analysis of Au+Au, 
Cu+Cu and $d$+Au data, the combinatorial background is estimated using an 
event-mixing technique.  The details of the method are given 
elsewhere~\cite{PPG016}.  In the ``no PID'' analysis, a significant residual 
background remains in the subtracted mass spectra because the mixed-event 
technique does not account for the abundant correlated pairs from other 
particle decays (K$_{s}^{0} \rightarrow \pi^{+}\pi^{-}$, 
$\Lambda \rightarrow p\pi^{-}$, $\rho \rightarrow \pi^{+}\pi^{-}$, 
$\omega \rightarrow \pi^{0}\pi^{+}\pi^{-}$ etc.).  In the ``one kaon PID'' 
analysis the residual background is considerably smaller~\cite{PPG099} 
while in the ``two kaon PID'' method the background is negligible.  Examples 
of subtracted mass spectra obtained in Au+Au collisions with the``two kaon 
PID'' and ``no PID'' techniques are shown in Fig.~\ref{fig:Minv}.  The S/B 
ratio depends on the collision system, the analysis technique, the $\phi$ 
transverse momentum and the centrality.  The typical ranges of the S/B 
values for each collision system and each analysis technique are summarized 
in Table~\ref{nevt}.

The total combinatorial background in $p$+$p$~\cite{PPG099}, as well as the 
residual background in $d$+Au, Cu+Cu and Au+Au analyses were estimated by 
fitting the mass spectra with the sum of a Breit-Wigner mass distribution 
function convolved with a Gaussian experimental mass resolution function to 
account for the $\phi$ signal, and a polynomial function to account for the 
background.  The typical experimental mass resolution for the $\phi$ meson 
was estimated to be $\sim$ 1~MeV/$c^{2}$ using Monte-Carlo studies based on 
the known momentum resolution of the tracking system and time resolution of 
the TOF.  To describe the background a second order polynomial was used in 
most analyses, except for the Au+Au ``no PID'' case where a third order 
polynomial was used.  Figure~\ref{fig:Minv} shows an example of the fits. 

The $\phi$ meson invariant yield in a given centrality and $p_T$ bin 
is obtained by:

\begin{equation}
\label{formula1}
\frac{1}{2 \pi p_T} \frac{d^{2}N}{dp_Tdy} =
\frac{N_{\phi}C_{\rm bias}} {2\pi p_TN_{\rm evt}
\epsilon_{\rm rec}\epsilon_{\rm embed}B_{\rm KK}\Delta p_{\rm
    T}\Delta y},
\end{equation}
where $N_{\rm evt}$ is the number of analyzed events in the centrality bin 
under consideration, $\epsilon_{\rm rec}$ corrects for the limited 
acceptance of the detector and for the $\phi$ meson reconstruction 
efficiency, $\epsilon_{\rm embed}$ accounts for the losses in 
reconstruction efficiency due to detector occupancy in heavy ion 
collisions, $B_{\rm KK}$ is the branching ratio for $\phi\rightarrow$ 
K$^+$K$^-$ in vacuum, $N_{\phi}$ is the raw $\phi$ yield measured in the 
given bin, $C_{\rm bias} = \epsilon_{\rm MB}^{\rm BBC} / 
\epsilon_{\phi}^{\rm BBC}$, where $\epsilon_{\rm MB}^{\rm BBC}$ and 
$\epsilon_{\phi}^{\rm BBC}$ are the beam-beam trigger efficiencies for 
minimum bias and $\phi$ events respectively.  This $C_{\rm bias}$ 
correction is equal to 0.69 for $p$+$p$~\cite{biaspp} and varies from 0.92 
to 0.85 as we go from peripheral to central $d$+Au 
collisions~\cite{biasdAu}.  In Au+Au and Cu+Cu collisions the minimum bias 
trigger is inefficient only for very peripheral collisions (centrality 
$>92.2\%$ for Au+Au and $>94\%$ for Cu+Cu).  For all other centralities, 
$0-92.2\%$ ($0-94\%$) for Au+Au (Cu+Cu), there is no trigger bias and 
$C_{\rm bias}=1$.  In $p$+$p$ the invariant differential cross section at 
midrapidity is related to invariant yield as $E \frac{d^{3}\sigma}{dp^{3}} 
= \sigma_{pp}^{inel} \times \frac{1}{2\pi p_T} \frac{d^{2}N}{dp_{\rm 
T}dy}$, where $\sigma_{inel}^{pp}$ = 42.2~mb.
 
The detector occupancy related loss $(1 - \epsilon_{\rm embed})$ is 
calculated by embedding simulated K$^+$K$^-$ pairs into real events.  It 
varies from 1\% in peripheral to 29\% (7\%) in the most central Au+Au 
(Cu+Cu) collisions without a significant $p_T$ dependence.

\begin{figure}[thb]
 \includegraphics[width=0.99\linewidth]{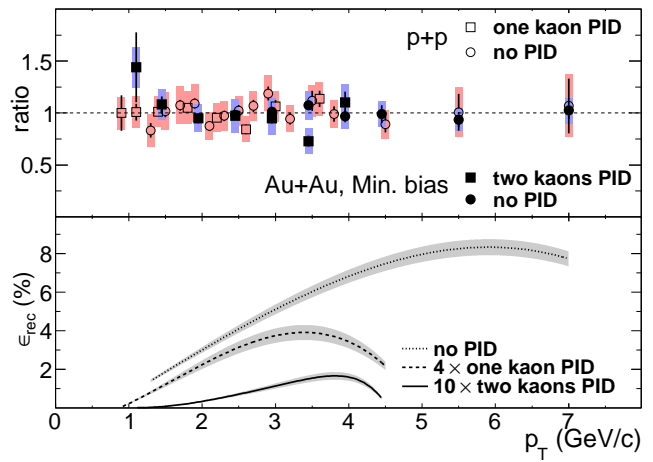} 
 \caption{\label{fig:techplot} (Color online) 
Top: Ratios of yields obtained with ``no PID'' and ``one kaon PID'' 
(in $p$+$p$ collisions), or ``no PID'' and ``two kaons PID'' 
(in Au+Au collisions) to fits to the combined spectra.  
Bottom: comparison of the acceptance and reconstruction efficiency 
for the three different analysis techniques.  
}
\end{figure}

The acceptance and reconstruction efficiency ($\epsilon_{\rm rec}$) of the 
$\phi$ meson, determined using a full GEANT simulation of the PHENIX 
detector, is shown in the bottom panel of Fig.~\ref{fig:techplot} for 
different analysis techniques.  There are very large differences, reaching 
more than one order of magnitude between the three cases.  In spite of 
that, the invariant yield spectra obtained from the different techniques 
are in good agreement as demonstrated in the top panel of 
Fig.~\ref{fig:techplot}, which shows the ratios of yields obtained 
with ``no PID'' or with ``one kaon PID'' (``no PID'' or ``two kaons PID'') 
techniques in $p$+$p$ (Au+Au) to a fit performed to the combined data sets.  
This agreement implies good control over the systematic uncertainties which 
are quite different in the three cases and provides confidence on the 
robustness of the experimental results.

The results from measurements at low $p_T$ using ``two kaons PID'' 
(in Au+Au collisions) and ``one kaon PID'' (in $p$+$p$ and Cu+Cu) are 
combined with the independent ``no PID'' measurements at intermediate and 
high $p_T$ to form the final $p_T$ spectra.  The measurement in 
$d$+Au is performed using the ``no PID'' technique only.  The invariant mass 
spectra obtained with ``one kaon PID'' or ``two kaon PID'' methods are 
subsamples of the ``no PID'' distribution.  Therefore results obtained with 
different methods can not be directly averaged.  In the final spectra the 
transition between different techniques occurs at $p_T=1.3~$GeV/$c$ 
in $p$+$p$, $p_T=2.2~$GeV/$c$ in Au+Au and at $p_T=3.2~$GeV/$c$ 
in Cu+Cu collisions in order to obtain the smallest combined statistical 
and systematical uncertainties for the points.

\begin{figure*}[thb]
  \includegraphics[width=0.99\linewidth]{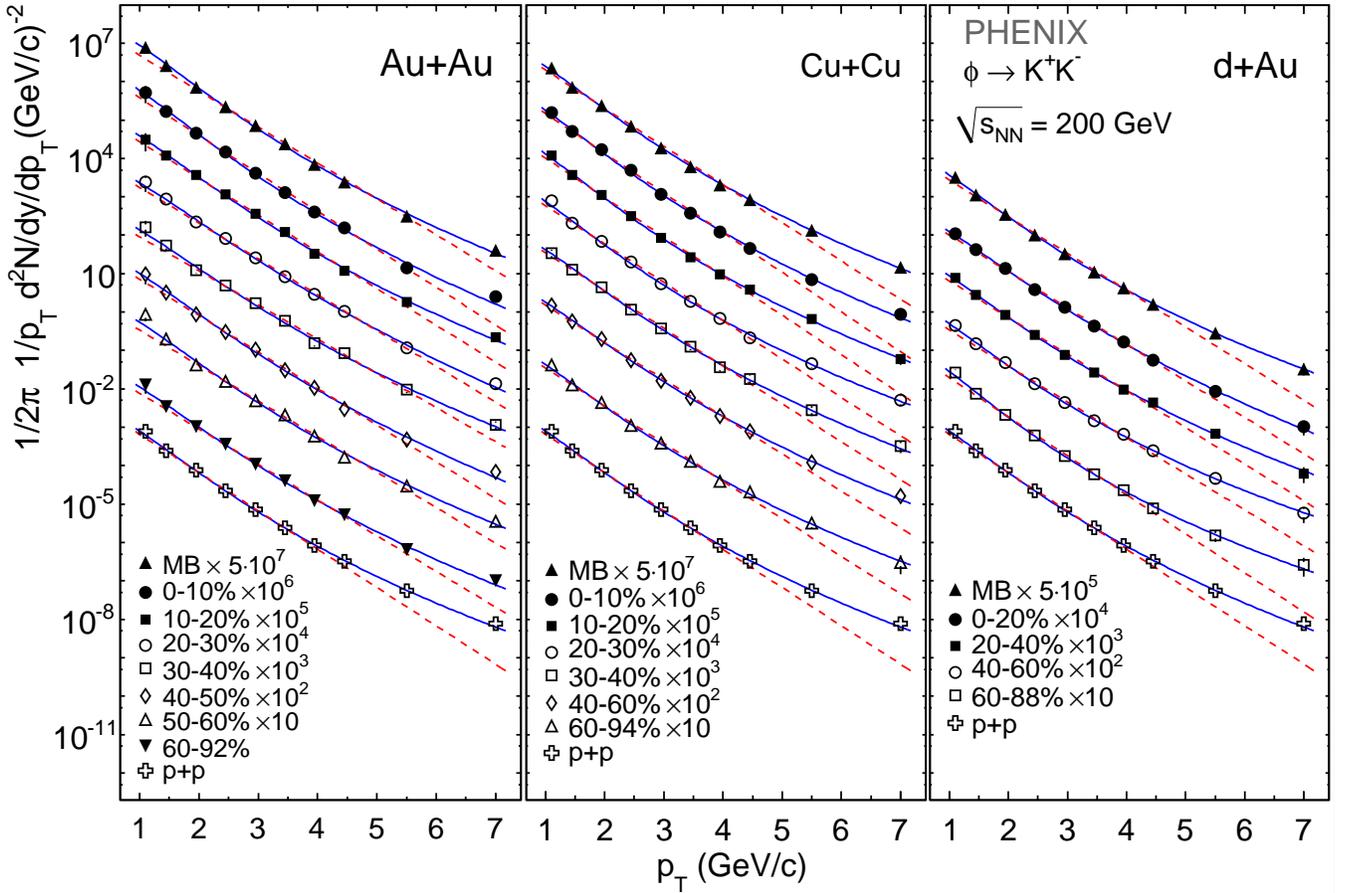}  
 \caption{\label{fig:spectra} (Color online) 
Invariant $p_T$ spectra of the $\phi$ meson for different centrality 
bins in Au+Au, Cu+Cu, $d$+Au and $p$+$p$ collisions at \sqsn=200~GeV.  The 
statistical and systematic uncertainties are smaller than the size of the 
symbols.  The spectra are fitted to exponential and 
Tsallis~\cite{Tsallis1988,Tsallis1999,Wilk2000} functions shown by the 
dashed and solid lines, respectively.
}
\end{figure*}

Systematic uncertainties on the $\phi$ invariant yield are grouped into 
three categories: type A (point-to-point uncorrelated), which can move each 
point independently; type B (point-to-point $p_T$-correlated), which 
can move points coherently, but not necessarily by the same relative 
amount; type C (global), which move all points by the same relative amount.  
The main contribution to the systematic errors of type A is the uncertainty 
in the raw yield extraction $N_{\phi}$ of 6--25\%.  Error of type B is 
dominated by uncertainties in reconstruction efficiency $\epsilon_{\rm 
rec}$ of 5--9\%, embedding corrections $\epsilon_{\rm embed}$ of 
1--7\% and momentum scale of 1--5\%.  The main contributions to the 
type C errors are the uncertainties in normalization for the $p$+$p$ 
($d$+Au) cross section equal to $9.7\%$ ($7.8\%$) and in branching ratio 
$B_{\rm KK}$ of $1.2\%$.

\section{Results and Discussion}

Figure~\ref{fig:spectra} shows the fully corrected $\phi$ invariant yield 
as a function of $p_T$ measured in $p$+$p$, $d$+Au, Cu+Cu and Au+Au 
collisions at \sqsn=200~GeV.  The spectra are scaled by arbitrary factors 
for clarity and fitted to exponential and 
Tsallis~\cite{Tsallis1988,Tsallis1999,Wilk2000} functions shown by the 
dashed and solid lines, respectively.  We used the Tsallis function adapted 
to the form \cite{PPG099}:
\begin{eqnarray}
\label{tsalis}
\frac{1}{2\pi}\frac{d^{2}N}{dydp_T} & = &
\frac{1}{2\pi}\frac{dN}{dy}\frac{(n-1)(n-2)}{(nT+m_{\phi}(n-1))(nT+m_{\phi})}
\nonumber  \\
& \times & \left(\frac{nT+m_T}{nT+m_{\phi}}\right)^{-n},
\end{eqnarray} 
where $\frac{dN}{dy}$, $n$ and $T$ are free parameters, 
$m_T = \sqrt{p^2_T + m^2_{\phi}}$, and $m_{\phi}$ is the mass of the 
$\phi$ meson.  The spectral shapes for all collision systems and 
centralities are well described by the Tsallis function, while the 
exponential fits underestimate the $\phi$ meson yields at high $p_T$ where 
the spectra begin to exhibit the power law behavior expected for particles 
produced in hard scattering processes.  For $p$+$p$ collisions the 
departure from exponential shape occurs at $\approx$4 Gev/$c$.  For all 
centralities in Au+Au collisions the departure occurs at somewhat larger 
$p_T$, which suggests a larger contribution of soft processes 
to the $\phi$ meson production up to 4--5~GeV/$c$.  Such behavior 
of the spectral shapes is in agreement with recombination
models~\cite{recombination2,Hwa:2002tu,recombination1,hadroproduction,Hwa2007,Hwa2004} 
predicting $p_T$ spectra for different hadronic species based on the 
number and flavor of constituent quarks.  At low transverse momentum, we do 
not observe a large change in the slopes of the spectra from central to 
peripheral collisions, supporting the expectation for smaller radial flow 
in $\phi$ mesons compared to other hadrons.  

The large $p_T$ reach of the results presented here allows for the 
study of medium-induced effects on $\phi$ meson production at intermediate 
and high $p_T$ using the nuclear modification factor:
\begin{equation}
\label{formula2}
  R_{\rm AB}(\pt)= dN_{\rm AB}/(\langle N_{\rm coll} \rangle
  \times dN_{pp}),
\end{equation}
where $dN_{\rm AB}$ ($dN_{pp}$) is the differential $\phi$ yield in 
nucleus-nucleus ($p$+$p$) collisions and $\langle N_{\rm coll} \rangle$ is 
the average number of nuclear collisions in the centrality bin under 
consideration~\cite{PPG014,PPG030,PPG084}.  The latter is determined solely 
by the density distribution of the nucleons in the nuclei A and B and by 
the impact parameter and is calculated using the Glauber 
formalism~\cite{GLAUBER}.  Deviation of $R_{\rm AB}$ from unity quantifies 
the degree of departure of the A+B yields from a superposition of 
incoherent nucleon-nucleon collisions.

\begin{figure}[thb]
\includegraphics[width=0.99\linewidth]{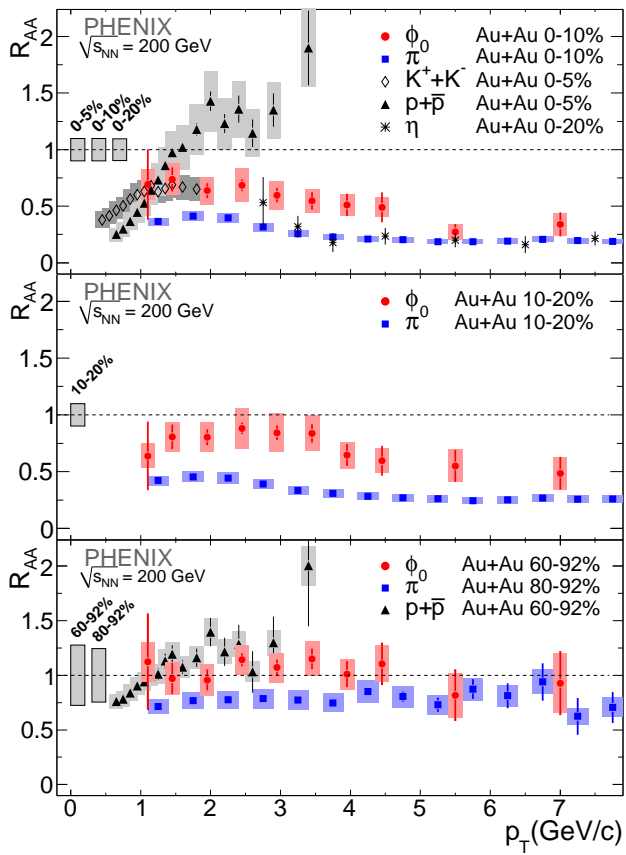} 
\caption{\label{fig:raa_auau} (Color online) 
Top: $R_{\rm AA}$ vs.  $p_T$ for $\phi$, $\pi^0$, $\eta$, 
(K$^+$+K$^-$) and ($\rm p+\bar{\rm p}$) in central Au+Au collisions.
Middle: $R_{\rm AA}$ vs.  $p_T$ for $\phi$ and $\pi^0$ in 10--20\% 
mid-central Au+Au collisions.
Bottom: $R_{\rm AA}$ vs.  $p_T$ for $\phi$, and $\rm p+\bar{\rm p}$ in 
60--92\% and for $\pi^0$ in 80--92\% peripheral Au+Au collisions.  
Values for (K$^+$+K$^-$), ($\rm p+\bar{\rm p}$), $\pi^0$ and $\eta$ are 
from~\cite{PPG030,PPG044,PPG080,PPG084,PPG051}.
The uncertainty in the determination of $\langle N_{\rm coll} \rangle$ is 
shown as a box on the left.  The global uncertainty of $\sim10\%$ related to 
the $p$+$p$ reference normalization is not shown.
}
\end{figure}

\begin{figure}[tb]
\includegraphics[width=0.99\linewidth]{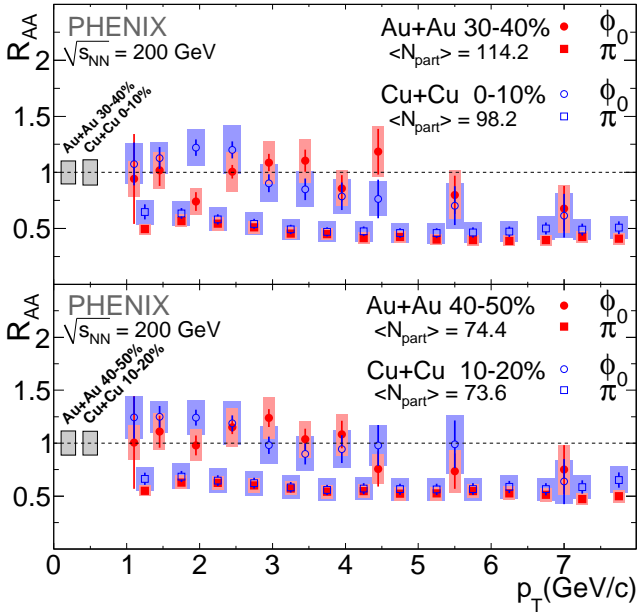} 
 \caption{\label{fig:raa_aucu} (Color online) 
Top: $R_{\rm AA}$ vs.  $p_T$ for $\phi$ and $\pi^0$ for 30--40\% 
centrality Au+Au and 0--10\% centrality Cu+Cu collisions.
Bottom: $R_{\rm AA}$ vs.  $p_T$ for $\phi$ and $\pi^0$ for 40--50\% 
centrality Au+Au and 10--20\% centrality Cu+Cu collisions.  Values for 
$\pi^0$ are from~\cite{PPG080,PPG084}.  The uncertainty in the determination 
of $\langle N_{\rm coll} \rangle$ is shown as a box on the left.  The global 
uncertainty of $\sim10\%$ related to the $p$+$p$ reference normalization is 
not shown.
}
\end{figure}

Figure~\ref{fig:raa_auau} shows a comparison of the R$_{\rm AA}$ for 
$\phi$ and $\pi^0$ from Ref.~\cite{PPG080}, proton and kaon from 
Ref.~\cite{PPG030} and $\eta$ from Ref.~\cite{PPG051}, all measured in 
Au+Au collisions at \sqsn=200~GeV.  The $\phi$ meson exhibits a different 
suppression pattern than that of lighter non-strange mesons and baryons.  
For central collisions (top panel) the $\phi$'s R$_{\rm AA}$ shows less 
suppression than $\pi^0$ and $\eta$ in the intermediate $p_T$ range 
of $2<p_T$~(GeV/$c$)$<5$.  At higher $p_T$ values, $p_{\rm T}>5$~GeV/$c$, 
the $\phi$'s R$_{\rm AA}$ approaches and becomes comparable 
to the $\pi^0$ and $\eta$ R$_{\rm AA}$.  These two features remain true for 
all centralities up to the most peripheral collisions as displayed in the 
bottom panel of Fig.~\ref{fig:raa_auau} (see also Fig.~\ref{fig:raa_aucu}).  
The panel shows that the $\pi^0$ is slightly suppressed (at the level of 
$\sim20\%$) in peripheral Au+Au collisions whereas the $\phi$ is not 
suppressed.  The kaon data cover only a very limited range at low $p_{\rm 
T}$ but in this range they seem to follow the R$_{\rm AA}$ trend of the 
$\phi$ better than that of the $\pi^0$ and $\eta$ for central Au+Au 
collisions.  The comparison with baryons, represented in 
Figure~\ref{fig:raa_auau} by the protons and anti-protons, shows a 
different pattern.  For central collisions, the protons show no suppression 
but rather an enhancement at $p_T>1.5$~GeV/$c$, whereas the $\phi$ 
mesons are suppressed.  This difference between $\phi$ mesons and protons 
gradually disappears with decreasing centrality and for the most 
peripheral collisions the R$_{\rm AA}$ of $\phi$ and (anti)protons are 
very similar as demonstrated in the bottom panel.

The results presented here are in agreement with the previous PHENIX 
results~\cite{PPG016}, which were based on the 2002 RHIC run, within the 
relatively larger uncertainties of the latter.  The use of different 
analysis techniques and the larger Au+Au data sample of the 2004 run 
resulted in a higher precision and a larger $p_T$ reach of R$_{\rm 
AA}$ that allowed to unveil the different behavior of the $\phi$ meson, 
i.e.  less suppression than $\pi^0$ but more suppression than baryons, in 
the intermediate $p_T$ range.  Our results differ from the ones 
recently published by the STAR 
Collaboration~\cite{STAR-phi2007,STAR-phi2009} which show that in Au+Au 
collisions R$_{\rm AA}$ is consistent with binary scaling in the 
intermediate $p_T$ region whereas R$_{\rm CP}$ shows considerable 
suppression.  This difference is traced down to the almost factor of two 
higher invariant $p_T$ yield in the STAR 
experiment~\cite{STAR-phi2007,STAR-phi2009} in Au+Au collisions, compared 
to our results presented in Fig.~\ref{fig:spectra}, whereas in $p+p$ both 
experiments are in reasonably good agreement.

\begin{figure}[tb]
  \includegraphics[width=0.9\linewidth]{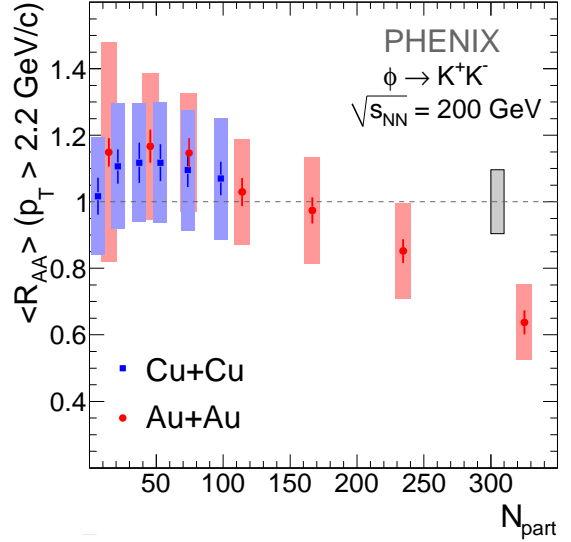} 
\caption{\label{fig:raa_int} (Color online) 
$R_{\rm AA}$ for $\phi$ integrated at $p_T > 2.2$~GeV/$c$ in Cu+Cu 
and Au+Au collisions vs.  $N_{\rm part}$.  The global uncertainty related to the 
$p+p$ reference normalization is shown as a box on the right.  }
\end{figure}

Figure~\ref{fig:raa_aucu} compares the R$_{\rm AA}$ of $\phi$ in Au+Au and 
Cu+Cu in two centrality bins which correspond approximately to the same 
number of participants in the two systems.  Figure~\ref{fig:raa_int} shows 
the $R_{\rm AA}$ of the $\phi$ integrated for $p_T > 2.2$~GeV/$c$ in 
Cu+Cu and Au+Au collisions versus $N_{\rm part}$.  Under these conditions, there 
is no difference in the R$_{\rm AA}$ of $\phi$ between the two systems 
indicating that the level of the suppression, when averaged over the 
azimuthal angle, scales with the average size of the nuclear overlap, 
regardless of the details of its shape.  This behavior has been observed in 
other measurements, such as the R$_{\rm AA}$ of the $\pi^0$.  The $\pi^0$ 
suppression data in Au+Au and Cu+Cu taken from Ref.~\cite{PPG080,PPG084} 
are also shown in Fig.~\ref{fig:raa_aucu} for comparison.  The similarity of 
the R$_{\rm AA}$ of $\phi$ in the two colliding systems implies that the 
features discussed previously for Au+Au in the context of 
Fig.~\ref{fig:raa_auau}, namely that the $\phi$ exhibits an intermediate 
suppression between pions and baryons, remain valid also in the Cu+Cu 
system.

Our data disfavor radial flow as the dominant source for the particle 
species dependence of the suppression factors at intermediate $p_T$, 
since the proton and $\phi$ R$_{\rm AA}$ factors differ by a factor of 
$\sim2$, in spite of their similar mass ($m_{\rm p} \simeq m_\phi$), 
whereas the kaon and $\phi$ show similar R$_{\rm AA}$ factors although 
their masses differ by almost a factor of two ($m_{\phi} \simeq 2m_{\rm 
K}$).

Recombination 
models~\cite{recombination1,hadroproduction,recombination2,Hwa:2002tu,Hwa2007,Hwa2004} 
qualitatively explain the larger yield of baryons compared to mesons at 
intermediate $p_T$ by the higher gain in $p_T$ which comes from 
recombination of three quarks for baryons rather than two quarks for 
mesons.  The same framework can be used to interpret the difference in 
suppression factors for $\pi^0$ and $\phi$ mesons.  
For $\pi^0$ production in the Hwa and Yang model~\cite{Hwa2004} 
the contribution from the recombination of thermal (T) and 
shower (S) partons becomes comparable to that of the recombination of TT 
partons already at $p_T\approx3$~GeV/$c$.  For the 
$\phi$ however, the strangeness enhancement feeds preferentially the 
thermal quarks.  Soft processes dominate over hard processes in a wider 
$p_T$ range and consequently the TT component remains dominant up to 
$p_T\approx6$~GeV/$c$ for the $\phi$ production~\cite{Hwa2007}.  The 
R$_{\rm AA}$ of $\phi$ becomes similar to that for $\pi^0$ at $p_{\rm 
T}>5-6$~GeV/$c$ where the contribution from fragmentation partons becomes 
significant for both particles.  It is interesting to note that the $\eta$ 
follows closely the $\pi^0$ in spite of its sizable ($\sim 50\%$) 
strangeness content~\cite{Uvarov2001}.

\begin{figure}[hb]
\includegraphics[width=0.99\linewidth]{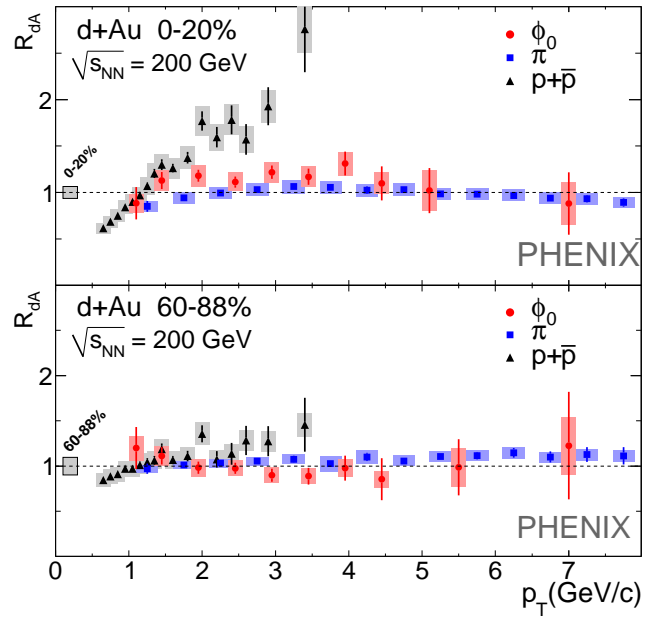} 
 \caption{\label{fig:rda} (Color online) 
Top: $R_{\rm dA}$ vs.  $p_T$ for $\phi$, $\pi^0$ and ($\rm p+\bar{\rm 
p}$) for 0--20\% centrality $d$+Au collisions.
Bottom: $R_{\rm dA}$ vs.  $p_T$ for $\phi$, $\pi^0$ and ($\rm 
p+\bar{\rm p}$) for 60--88\% peripheral $d$+Au collisions.  Values for 
(K$^+$+K$^-$) and ($\rm p+\bar{\rm p}$) and $\pi^0$ are 
from~\cite{PPG030,PPG044}.  The uncertainty in the determination of $\langle 
N_{\rm coll} \rangle$ is shown as a box on the left.  The global uncertainty 
of $\sim10\%$ related to the $p$+$p$ reference normalization is not shown.
}
\end{figure}

\clearpage

Cold nuclear matter effects can also contribute to the differences in 
hadron suppression factors in A+A collisions.  Figure~\ref{fig:rda}  
compares the R$_{\rm dA}$ for $\phi$ and $\pi^0$ from Ref.~\cite{PPG044}, 
and protons from Ref.~\cite{PPG030} for central (top panel) and peripheral 
(bottom panel) $d$+Au collisions.  For both centralities, the R$_{\rm dA}$ 
for $\phi$ and $\pi^0$ are similar indicating that cold nuclear effects are 
not responsible for the differences between $\phi$ and $\pi^0$ seen in 
Au+Au and Cu+Cu collisions.  The proton's R$_{\rm dA}$ exhibits an 
enhancement for $p_T$=2--4~GeV/$c$, usually associated with the Cronin 
effect~\cite{Cronin1,Cronin2,Cronin3,Cronin4,Cronin5,Accardi2004}, 
whereas the R$_{\rm dA}$ for $\phi$ indicates little or no enhancement.  The 
lack of Cronin enhancement is also seen in the $\pi^0$ data~\cite{PPG044} 
shown in Fig.~\ref{fig:rda} and has also been observed for other mesons in 
central and mid-central $d$+Au collisions at 
\sqsn=200~GeV~\cite{PPG030,STAR_dAu1,STAR_dAu2}.

\section{Summary and Conclusions}

We have measured $\phi$ meson production at midrapidity via the K$^+$K$^-$ 
decay channel in $p$+$p$, $d$+Au, Cu+Cu and Au+Au collisions at 
\sqsn=200~GeV.  Invariant $p_T$ spectra and nuclear modification 
factors have been presented over the $p_T$ range of 
$1<p_T<7$ GeV/$c$ for different centralities.

The $\phi$ meson exhibits a different suppression pattern compared to 
lighter mesons ($\pi^0$ and $\eta$) and baryons (protons and anti-protons) 
in heavy ion collisions.  For all centralities, the $\phi$ meson is less 
suppressed than $\pi^0$ and $\eta$ in the intermediate $p_T$ range 
(2--5~GeV/$c$) whereas at higher $p_T$, $\phi$, $\pi^0$ and $\eta$ show 
similar suppression values.  The available kaon R$_{\rm AA}$ data seem to 
follow the R$_{\rm AA}$ trend of the $\phi$.  The comparison with baryons 
shows that in central Au+Au collisions the latter are enhanced with 
respect to binary scaling whereas the $\phi$ meson is suppressed, but this 
difference gradually disappears with decreasing centrality and for 
peripheral collisions the baryons and the $\phi$ meson have very similar 
R$_{\rm AA}$ values consistent with binary scaling.

The same features are observed in Cu+Cu collisions between the $\phi$ and 
$\pi^0$.  The $\phi$ meson invariant $p_T$ spectra in Au+Au and Cu+Cu 
collisions for similar \Np values exhibit similar shape and yield over the 
entire $p_T$ range of the measurement within the statistical and 
systematic uncertainties.  This indicates that the production and 
suppression of the $\phi$ meson, when averaged over the azimuthal angle, 
scales with the average size of the nuclear overlap region, regardless of 
the details of its shape.

Cold nuclear effects cannot account for the observed differences.  For all 
centralities, the $\phi$'s R$_{\rm dA}$ in $d$+Au collisions is consistent 
with binary scaling in agreement with other mesons.  No meson species 
dependence is observed in R$_{\rm dA}$ within uncertainties.

The observed features at intermediate $p_T$ in Au+Au and Cu+Cu 
collisions are qualitatively consistent with quark recombination 
models~\cite{recombination1,hadroproduction,recombination2,Hwa:2002tu,Hwa2007,Hwa2004}, 
which are also supported by $\phi$ elliptic flow 
measurements~\cite{PPG073,STAR-phi2007}.  The systematic set of measurements 
presented here provides further constraints to these models.  The similarity 
between the suppression patterns of different mesons at high $p_T$ 
favors the production of these mesons via jet fragmentation outside the hot 
and dense medium created in the collision.  Complementary jet correlation 
measurements involving $\phi$ mesons as a trigger as well as extension of 
the kaon data to higher $p_T$ would be desirable to provide further 
insight into the $\phi$ meson production mechanism.
 
\section*{ACKNOWLEDGEMENTS}


We thank the staff of the Collider-Accelerator and Physics
Departments at Brookhaven National Laboratory and the staff of
the other PHENIX participating institutions for their vital
contributions.  We acknowledge support from the 
Office of Nuclear Physics in the
Office of Science of the Department of Energy, the
National Science Foundation, Abilene Christian University
Research Council, Research Foundation of SUNY, and Dean of the
College of Arts and Sciences, Vanderbilt University (U.S.A),
Ministry of Education, Culture, Sports, Science, and Technology
and the Japan Society for the Promotion of Science (Japan),
Conselho Nacional de Desenvolvimento Cient\'{\i}fico e
Tecnol{\'o}gico and Funda\c c{\~a}o de Amparo {\`a} Pesquisa do
Estado de S{\~a}o Paulo (Brazil),
Natural Science Foundation of China (People's Republic of China),
Ministry of Education, Youth and Sports (Czech Republic),
Centre National de la Recherche Scientifique, Commissariat
{\`a} l'{\'E}nergie Atomique, and Institut National de Physique
Nucl{\'e}aire et de Physique des Particules (France),
Ministry of Industry, Science and Tekhnologies,
Bundesministerium f\"ur Bildung und Forschung, Deutscher
Akademischer Austausch Dienst, and Alexander von Humboldt Stiftung (Germany),
Hungarian National Science Fund, OTKA (Hungary), 
Department of Atomic Energy and Department of Science and Technology (India), 
Israel Science Foundation (Israel), 
National Research Foundation (Korea),
Ministry of Education and Science, Russia Academy of Sciences,
Federal Agency of Atomic Energy (Russia),
VR and the Wallenberg Foundation (Sweden), 
the U.S. Civilian Research and Development Foundation for the
Independent States of the Former Soviet Union, 
the US-Hungarian Fulbright Foundation for Educational Exchange,
and the US-Israel Binational Science Foundation.



\begin{thebibliography}{59}
\expandafter\ifx\csname natexlab\endcsname\relax\def\natexlab#1{#1}\fi
\expandafter\ifx\csname bibnamefont\endcsname\relax
  \def\bibnamefont#1{#1}\fi
\expandafter\ifx\csname bibfnamefont\endcsname\relax
  \def\bibfnamefont#1{#1}\fi
\expandafter\ifx\csname citenamefont\endcsname\relax
  \def\citenamefont#1{#1}\fi
\expandafter\ifx\csname url\endcsname\relax
  \def\url#1{\texttt{#1}}\fi
\expandafter\ifx\csname urlprefix\endcsname\relax\def\urlprefix{URL }\fi
\providecommand{\bibinfo}[2]{#2}
\providecommand{\eprint}[2][]{\url{#2}}

\bibitem[{\citenamefont{Braun-Munzinger et~al.}(2003)}]{BraunMunzinger:2003zd}
\bibinfo{author}{\bibfnamefont{P.}~\bibnamefont{Braun-Munzinger}}
  \bibnamefont{et~al.}, \bibinfo{journal}{Quark-gluon plasma}
  \textbf{\bibinfo{volume}{3}}, \bibinfo{pages}{491} (\bibinfo{year}{2003}).

\bibitem[{\citenamefont{Adcox et~al.}(2005)}]{PHENIX_white}
\bibinfo{author}{\bibfnamefont{K.}~\bibnamefont{Adcox}} \bibnamefont{et~al.}
  (\bibinfo{collaboration}{PHENIX Collaboration}), \bibinfo{journal}{Nucl.
  Phys.} \textbf{\bibinfo{volume}{A757}}, \bibinfo{pages}{184}
  (\bibinfo{year}{2005}).

\bibitem[{\citenamefont{Kolb and Heinz}(2003)}]{Kolb:2003dz}
\bibinfo{author}{\bibfnamefont{P.~F.} \bibnamefont{Kolb}} \bibnamefont{and}
  \bibinfo{author}{\bibfnamefont{U.~W.} \bibnamefont{Heinz}},
  \bibinfo{journal}{Quark-gluon plasma} \textbf{\bibinfo{volume}{3}},
  \bibinfo{pages}{634} (\bibinfo{year}{2003}).

\bibitem[{\citenamefont{Muller and Nagle}(2006)}]{Jamie}
\bibinfo{author}{\bibfnamefont{B.}~\bibnamefont{Muller}} \bibnamefont{and}
  \bibinfo{author}{\bibfnamefont{J.~L.} \bibnamefont{Nagle}},
  \bibinfo{journal}{Ann. Rev. Nucl. Part. Sci.} \textbf{\bibinfo{volume}{56}},
  \bibinfo{pages}{93} (\bibinfo{year}{2006}).

\bibitem[{\citenamefont{Huovinen and Ruuskanen}(2006)}]{Huovinen}
\bibinfo{author}{\bibfnamefont{P.}~\bibnamefont{Huovinen}} \bibnamefont{and}
  \bibinfo{author}{\bibfnamefont{P.}~\bibnamefont{Ruuskanen}},
  \bibinfo{journal}{Ann. Rev. Nucl. Part. Sci.} \textbf{\bibinfo{volume}{56}},
  \bibinfo{pages}{163} (\bibinfo{year}{2006}).

\bibitem[{\citenamefont{Gyulassy et~al.}(1990)}]{jetquenching}
\bibinfo{author}{\bibfnamefont{M.}~\bibnamefont{Gyulassy}}
  \bibnamefont{et~al.}, \bibinfo{journal}{Phys. Lett.}
  \textbf{\bibinfo{volume}{B243}}, \bibinfo{pages}{432 }
  (\bibinfo{year}{1990}).

\bibitem[{\citenamefont{Baier et~al.}(1997)}]{Baier1997}
\bibinfo{author}{\bibfnamefont{R.}~\bibnamefont{Baier}} \bibnamefont{et~al.},
  \bibinfo{journal}{Nucl. Phys.} \textbf{\bibinfo{volume}{B484}},
  \bibinfo{pages}{265} (\bibinfo{year}{1997}).

\bibitem[{\citenamefont{Gyulassy et~al.}(1994)}]{Gyulassy1994}
\bibinfo{author}{\bibfnamefont{M.}~\bibnamefont{Gyulassy}}
  \bibnamefont{et~al.}, \bibinfo{journal}{Nucl. Phys.}
  \textbf{\bibinfo{volume}{B420}}, \bibinfo{pages}{583} (\bibinfo{year}{1994}).

\bibitem[{\citenamefont{d\'Enterria}(2009)}]{jet_quench}
\bibinfo{author}{\bibfnamefont{D.}~\bibnamefont{d\'Enterria}}
  (\bibinfo{year}{2009}), \bibinfo{note}{arXiv:0902.2011/nucl-ex}.

\bibitem[{\citenamefont{Adcox et~al.}(2001)}]{PPG003}
\bibinfo{author}{\bibfnamefont{K.}~\bibnamefont{Adcox}} \bibnamefont{et~al.}
  (\bibinfo{collaboration}{PHENIX Collaboration}), \bibinfo{journal}{Phys. Rev.
  Lett.} \textbf{\bibinfo{volume}{88}}, \bibinfo{pages}{022301}
  (\bibinfo{year}{2001}).

\bibitem[{\citenamefont{Adler et~al.}(2003{\natexlab{a}})}]{PPG014}
\bibinfo{author}{\bibfnamefont{S.~S.} \bibnamefont{Adler}} \bibnamefont{et~al.}
  (\bibinfo{collaboration}{PHENIX Collaboration}), \bibinfo{journal}{Phys. Rev.
  Lett.} \textbf{\bibinfo{volume}{91}}, \bibinfo{pages}{072301}
  (\bibinfo{year}{2003}{\natexlab{a}}).

\bibitem[{\citenamefont{Adare et~al.}(2008{\natexlab{a}})}]{PPG084}
\bibinfo{author}{\bibfnamefont{A.}~\bibnamefont{Adare}} \bibnamefont{et~al.}
  (\bibinfo{collaboration}{PHENIX Collaboration}), \bibinfo{journal}{Phys. Rev.
  Lett.} \textbf{\bibinfo{volume}{101}}, \bibinfo{pages}{162301}
  (\bibinfo{year}{2008}{\natexlab{a}}).

\bibitem[{\citenamefont{Adler et~al.}(2005{\natexlab{a}})}]{PPG042}
\bibinfo{author}{\bibfnamefont{S.~S.} \bibnamefont{Adler}} \bibnamefont{et~al.}
  (\bibinfo{collaboration}{PHENIX Collaboration}), \bibinfo{journal}{Phys. Rev.
  Lett.} \textbf{\bibinfo{volume}{94}}, \bibinfo{pages}{232301}
  (\bibinfo{year}{2005}{\natexlab{a}}).

\bibitem[{\citenamefont{Adare et~al.}(2007{\natexlab{a}})}]{heavy}
\bibinfo{author}{\bibfnamefont{A.}~\bibnamefont{Adare}} \bibnamefont{et~al.}
  (\bibinfo{collaboration}{PHENIX Collaboration}), \bibinfo{journal}{Phys. Rev.
  Lett.} \textbf{\bibinfo{volume}{98}}, \bibinfo{pages}{172301}
  (\bibinfo{year}{2007}{\natexlab{a}}).

\bibitem[{\citenamefont{Adler et~al.}(2003{\natexlab{b}})}]{PPG015}
\bibinfo{author}{\bibfnamefont{S.~S.} \bibnamefont{Adler}} \bibnamefont{et~al.}
  (\bibinfo{collaboration}{PHENIX Collaboration}), \bibinfo{journal}{Phys. Rev.
  Lett.} \textbf{\bibinfo{volume}{91}}, \bibinfo{pages}{172301}
  (\bibinfo{year}{2003}{\natexlab{b}}).

\bibitem[{\citenamefont{Adams et~al.}(2004)}]{K_Lambda_STAR}
\bibinfo{author}{\bibfnamefont{J.}~\bibnamefont{Adams}} \bibnamefont{et~al.}
  (\bibinfo{collaboration}{STAR Collaboration}), \bibinfo{journal}{Phys. Rev.
  Lett.} \textbf{\bibinfo{volume}{92}}, \bibinfo{pages}{052302}
  (\bibinfo{year}{2004}).

\bibitem[{\citenamefont{Adams et~al.}(2005{\natexlab{a}})}]{Kstar_levy}
\bibinfo{author}{\bibfnamefont{J.}~\bibnamefont{Adams}} \bibnamefont{et~al.}
  (\bibinfo{collaboration}{STAR Collaboration}), \bibinfo{journal}{Phys. Rev.
  C} \textbf{\bibinfo{volume}{71}}, \bibinfo{pages}{064902}
  (\bibinfo{year}{2005}{\natexlab{a}}).

\bibitem[{\citenamefont{Adler et~al.}(2005{\natexlab{b}})}]{Adler2005c}
\bibinfo{author}{\bibfnamefont{S.~S.} \bibnamefont{Adler}} \bibnamefont{et~al.}
  (\bibinfo{collaboration}{PHENIX Collaboration}), \bibinfo{journal}{Phys. Rev.
  C} \textbf{\bibinfo{volume}{71}}, \bibinfo{pages}{051902}
  (\bibinfo{year}{2005}{\natexlab{b}}).

\bibitem[{\citenamefont{Adare et~al.}(2007{\natexlab{b}})}]{PPG072}
\bibinfo{author}{\bibfnamefont{A.}~\bibnamefont{Adare}} \bibnamefont{et~al.}
  (\bibinfo{collaboration}{PHENIX Collaboration}), \bibinfo{journal}{Phys.
  Lett.} \textbf{\bibinfo{volume}{B649}}, \bibinfo{pages}{359}
  (\bibinfo{year}{2007}{\natexlab{b}}).

\bibitem[{\citenamefont{Hirano and Nara}(2004)}]{hydro2}
\bibinfo{author}{\bibfnamefont{T.}~\bibnamefont{Hirano}} \bibnamefont{and}
  \bibinfo{author}{\bibfnamefont{Y.}~\bibnamefont{Nara}},
  \bibinfo{journal}{Phys. Rev.} \textbf{\bibinfo{volume}{C69}},
  \bibinfo{pages}{034908} (\bibinfo{year}{2004}).

\bibitem[{\citenamefont{Heinz et~al.}(2002)}]{hydro4}
\bibinfo{author}{\bibfnamefont{U.~W.} \bibnamefont{Heinz}}
  \bibnamefont{et~al.}, \bibinfo{journal}{Nucl. Phys.}
  \textbf{\bibinfo{volume}{A702}}, \bibinfo{pages}{269} (\bibinfo{year}{2002}).

\bibitem[{\citenamefont{Greco et~al.}(2003)\citenamefont{Greco, Ko, and
  Levai}}]{recombination2}
\bibinfo{author}{\bibfnamefont{V.}~\bibnamefont{Greco}},
  \bibinfo{author}{\bibfnamefont{C.~M.} \bibnamefont{Ko}}, \bibnamefont{and}
  \bibinfo{author}{\bibfnamefont{P.}~\bibnamefont{Levai}},
  \bibinfo{journal}{Phys. Rev. Lett.} \textbf{\bibinfo{volume}{90}},
  \bibinfo{pages}{202302} (\bibinfo{year}{2003}).

\bibitem[{\citenamefont{Hwa and Yang}(2003)}]{Hwa:2002tu}
\bibinfo{author}{\bibfnamefont{R.~C.} \bibnamefont{Hwa}} \bibnamefont{and}
  \bibinfo{author}{\bibfnamefont{C.~B.} \bibnamefont{Yang}},
  \bibinfo{journal}{Phys. Rev.} \textbf{\bibinfo{volume}{C67}},
  \bibinfo{pages}{034902} (\bibinfo{year}{2003}).

\bibitem[{\citenamefont{Fries et~al.}(2003{\natexlab{a}})\citenamefont{Fries,
  Muller, Nonaka, and Bass}}]{recombination1}
\bibinfo{author}{\bibfnamefont{R.~J.} \bibnamefont{Fries}},
  \bibinfo{author}{\bibfnamefont{B.}~\bibnamefont{Muller}},
  \bibinfo{author}{\bibfnamefont{C.}~\bibnamefont{Nonaka}}, \bibnamefont{and}
  \bibinfo{author}{\bibfnamefont{S.~A.} \bibnamefont{Bass}},
  \bibinfo{journal}{Phys. Rev. Lett.} \textbf{\bibinfo{volume}{90}},
  \bibinfo{pages}{202303} (\bibinfo{year}{2003}{\natexlab{a}}).

\bibitem[{\citenamefont{Vitev and Gyulassy}(2002)}]{Vitev2002}
\bibinfo{author}{\bibfnamefont{I.}~\bibnamefont{Vitev}} \bibnamefont{and}
  \bibinfo{author}{\bibfnamefont{M.}~\bibnamefont{Gyulassy}},
  \bibinfo{journal}{Phys. Rev. C} \textbf{\bibinfo{volume}{65}},
  \bibinfo{pages}{041902} (\bibinfo{year}{2002}).

\bibitem[{\citenamefont{Brodsky et~al.}(2008)}]{brodsky}
\bibinfo{author}{\bibfnamefont{S.~J.} \bibnamefont{Brodsky}}
  \bibnamefont{et~al.}, \bibinfo{journal}{Phys. Lett.}
  \textbf{\bibinfo{volume}{B668}}, \bibinfo{pages}{111} (\bibinfo{year}{2008}).

\bibitem[{\citenamefont{Shor}(1985)}]{SomeReviewPaper}
\bibinfo{author}{\bibfnamefont{A.}~\bibnamefont{Shor}}, \bibinfo{journal}{Phys.
  Rev. Lett.} \textbf{\bibinfo{volume}{54}}, \bibinfo{pages}{1122}
  (\bibinfo{year}{1985}).

\bibitem[{\citenamefont{Adler et~al.}(2005{\natexlab{c}})}]{PPG016}
\bibinfo{author}{\bibfnamefont{S.~S.} \bibnamefont{Adler}} \bibnamefont{et~al.}
  (\bibinfo{collaboration}{PHENIX Collaboration}), \bibinfo{journal}{Phys. Rev.
  C} \textbf{\bibinfo{volume}{72}}, \bibinfo{pages}{014903}
  (\bibinfo{year}{2005}{\natexlab{c}}).

\bibitem[{\citenamefont{Abelev et~al.}(2007)}]{STAR-phi2007}
\bibinfo{author}{\bibfnamefont{B.}~\bibnamefont{Abelev}} \bibnamefont{et~al.}
  (\bibinfo{collaboration}{STAR Collaboration}), \bibinfo{journal}{Phys. Rev.
  Lett.} \textbf{\bibinfo{volume}{99}}, \bibinfo{pages}{112301}
  (\bibinfo{year}{2007}).

\bibitem[{\citenamefont{Abelev et~al.}(2009)}]{STAR-phi2009}
\bibinfo{author}{\bibfnamefont{B.}~\bibnamefont{Abelev}} \bibnamefont{et~al.}
  (\bibinfo{collaboration}{STAR Collaboration}), \bibinfo{journal}{Phys. Lett.}
  \textbf{\bibinfo{volume}{B673}}, \bibinfo{pages}{183} (\bibinfo{year}{2009}).

\bibitem[{\citenamefont{Afanasiev et~al.}(2007)}]{PPG073}
\bibinfo{author}{\bibfnamefont{S.}~\bibnamefont{Afanasiev}}
  \bibnamefont{et~al.} (\bibinfo{collaboration}{PHENIX Collaboration}),
  \bibinfo{journal}{Phys. Rev. Lett.} \textbf{\bibinfo{volume}{99}},
  \bibinfo{pages}{052301} (\bibinfo{year}{2007}).

\bibitem[{\citenamefont{Ko and Seibert}(1994)}]{phi_cross2}
\bibinfo{author}{\bibfnamefont{C.~M.} \bibnamefont{Ko}} \bibnamefont{and}
  \bibinfo{author}{\bibfnamefont{D.}~\bibnamefont{Seibert}},
  \bibinfo{journal}{Phys. Rev. C} \textbf{\bibinfo{volume}{49}},
  \bibinfo{pages}{2198} (\bibinfo{year}{1994}).

\bibitem[{\citenamefont{Adler et~al.}(2006{\natexlab{a}})}]{PPG030}
\bibinfo{author}{\bibfnamefont{S.~S.} \bibnamefont{Adler}} \bibnamefont{et~al.}
  (\bibinfo{collaboration}{PHENIX Collaboration}), \bibinfo{journal}{Phys. Rev.
  C} \textbf{\bibinfo{volume}{74}}, \bibinfo{pages}{024904}
  (\bibinfo{year}{2006}{\natexlab{a}}).

\bibitem[{\citenamefont{Hwa and Yang}(2004{\natexlab{a}})}]{reco}
\bibinfo{author}{\bibfnamefont{R.~C.} \bibnamefont{Hwa}} \bibnamefont{and}
  \bibinfo{author}{\bibfnamefont{C.~B.} \bibnamefont{Yang}},
  \bibinfo{journal}{Phys. Rev. Lett} \textbf{\bibinfo{volume}{93}},
  \bibinfo{pages}{082302} (\bibinfo{year}{2004}{\natexlab{a}}).

\bibitem[{\citenamefont{Adcox et~al.}(2003)}]{PHENIX}
\bibinfo{author}{\bibfnamefont{K.}~\bibnamefont{Adcox}} \bibnamefont{et~al.}
  (\bibinfo{collaboration}{PHENIX Collaboration}), \bibinfo{journal}{Nucl.
  Instrum. Meth.} \textbf{\bibinfo{volume}{A499}}, \bibinfo{pages}{469}
  (\bibinfo{year}{2003}).

\bibitem[{\citenamefont{Adcox et~al.}(2004)}]{centralityAuAu}
\bibinfo{author}{\bibfnamefont{K.}~\bibnamefont{Adcox}} \bibnamefont{et~al.}
  (\bibinfo{collaboration}{PHENIX Collaboration}), \bibinfo{journal}{Phys. Rev.
  C} \textbf{\bibinfo{volume}{69}}, \bibinfo{pages}{024904}
  (\bibinfo{year}{2004}).

\bibitem[{\citenamefont{Adler et~al.}(2005{\natexlab{d}})}]{centralitydAu}
\bibinfo{author}{\bibfnamefont{S.~S.} \bibnamefont{Adler}} \bibnamefont{et~al.}
  (\bibinfo{collaboration}{PHENIX Collaboration}), \bibinfo{journal}{Phys. Rev.
  Lett.} \textbf{\bibinfo{volume}{94}}, \bibinfo{pages}{082302}
  (\bibinfo{year}{2005}{\natexlab{d}}).

\bibitem[{\citenamefont{Adler et~al.}()}]{PPG099}
\bibinfo{author}{\bibfnamefont{S.~S.} \bibnamefont{Adler}} \bibnamefont{et~al.}
  (\bibinfo{collaboration}{PHENIX Collaboration}), \bibinfo{note}{(to be
  published)}.

\bibitem[{\citenamefont{Adler et~al.}(2007{\natexlab{a}})}]{biaspp}
\bibinfo{author}{\bibfnamefont{S.~S.} \bibnamefont{Adler}} \bibnamefont{et~al.}
  (\bibinfo{collaboration}{PHENIX Collaboration}), \bibinfo{journal}{Phys. Rev.
  Lett.} \textbf{\bibinfo{volume}{98}}, \bibinfo{pages}{172302}
  (\bibinfo{year}{2007}{\natexlab{a}}).

\bibitem[{\citenamefont{Adler et~al.}(2006{\natexlab{b}})}]{biasdAu}
\bibinfo{author}{\bibfnamefont{S.~S.} \bibnamefont{Adler}} \bibnamefont{et~al.}
  (\bibinfo{collaboration}{PHENIX Collaboration}), \bibinfo{journal}{Phys. Rev.
  Lett.} \textbf{\bibinfo{volume}{96}}, \bibinfo{pages}{012304}
  (\bibinfo{year}{2006}{\natexlab{b}}).

\bibitem[{\citenamefont{Tsallis}(1988)}]{Tsallis1988}
\bibinfo{author}{\bibfnamefont{C.}~\bibnamefont{Tsallis}}, \bibinfo{journal}{J.
  Stat. Phys.} \textbf{\bibinfo{volume}{52}}, \bibinfo{pages}{479}
  (\bibinfo{year}{1988}), \eprint{hep-ph/0804.4608}.

\bibitem[{\citenamefont{Tsallis}(1999)}]{Tsallis1999}
\bibinfo{author}{\bibfnamefont{C.}~\bibnamefont{Tsallis}},
  \bibinfo{journal}{Braz. J. Phys.} \textbf{\bibinfo{volume}{29}},
  \bibinfo{pages}{1} (\bibinfo{year}{1999}).

\bibitem[{\citenamefont{Wilk and Wlodarczyk}(2000)}]{Wilk2000}
\bibinfo{author}{\bibfnamefont{G.}~\bibnamefont{Wilk}} \bibnamefont{and}
  \bibinfo{author}{\bibfnamefont{Z.}~\bibnamefont{Wlodarczyk}},
  \bibinfo{journal}{Phys. Rev. Lett.} \textbf{\bibinfo{volume}{84}},
  \bibinfo{pages}{2770} (\bibinfo{year}{2000}), \eprint{hep-ph/9908459}.

\bibitem[{\citenamefont{Fries et~al.}(2003{\natexlab{b}})\citenamefont{Fries,
  Muller, Nonaka, and Bass}}]{hadroproduction}
\bibinfo{author}{\bibfnamefont{R.~J.} \bibnamefont{Fries}},
  \bibinfo{author}{\bibfnamefont{B.}~\bibnamefont{Muller}},
  \bibinfo{author}{\bibfnamefont{C.}~\bibnamefont{Nonaka}}, \bibnamefont{and}
  \bibinfo{author}{\bibfnamefont{S.~A.} \bibnamefont{Bass}},
  \bibinfo{journal}{Phys. Rev. C} \textbf{\bibinfo{volume}{68}},
  \bibinfo{pages}{044902} (\bibinfo{year}{2003}{\natexlab{b}}).

\bibitem[{\citenamefont{Hwa and Yang}(2006)}]{Hwa2007}
\bibinfo{author}{\bibfnamefont{R.~C.} \bibnamefont{Hwa}} \bibnamefont{and}
  \bibinfo{author}{\bibfnamefont{C.}~\bibnamefont{Yang}}
  (\bibinfo{year}{2006}), \bibinfo{note}{arXiv: nucl-th/0602024}.

\bibitem[{\citenamefont{Hwa and Yang}(2004{\natexlab{b}})}]{Hwa2004}
\bibinfo{author}{\bibfnamefont{R.~C.} \bibnamefont{Hwa}} \bibnamefont{and}
  \bibinfo{author}{\bibfnamefont{C.~B.} \bibnamefont{Yang}},
  \bibinfo{journal}{Phys. Rev. C} \textbf{\bibinfo{volume}{70}},
  \bibinfo{pages}{024905} (\bibinfo{year}{2004}{\natexlab{b}}),
  \eprint{nucl-th/0401001}.

\bibitem[{\citenamefont{Miller et~al.}(2007)}]{GLAUBER}
\bibinfo{author}{\bibfnamefont{M.~L.} \bibnamefont{Miller}}
  \bibnamefont{et~al.}, \bibinfo{journal}{Ann. Rev. Nucl. Part. Sci.}
  \textbf{\bibinfo{volume}{57}}, \bibinfo{pages}{205} (\bibinfo{year}{2007}),
  \eprint{nucl-ex/0701025}.

\bibitem[{\citenamefont{Adler et~al.}(2007{\natexlab{b}})}]{PPG044}
\bibinfo{author}{\bibfnamefont{S.~S.} \bibnamefont{Adler}} \bibnamefont{et~al.}
  (\bibinfo{collaboration}{PHENIX Collaboration}), \bibinfo{journal}{Phys. Rev.
  Lett.} \textbf{\bibinfo{volume}{98}}, \bibinfo{pages}{172302}
  (\bibinfo{year}{2007}{\natexlab{b}}).

\bibitem[{\citenamefont{Adare et~al.}(2008{\natexlab{b}})}]{PPG080}
\bibinfo{author}{\bibfnamefont{A.}~\bibnamefont{Adare}} \bibnamefont{et~al.}
  (\bibinfo{collaboration}{PHENIX Collaboration}), \bibinfo{journal}{Phys. Rev.
  Lett.} \textbf{\bibinfo{volume}{101}}, \bibinfo{pages}{232301}
  (\bibinfo{year}{2008}{\natexlab{b}}).

\bibitem[{\citenamefont{Adler et~al.}(2006{\natexlab{c}})}]{PPG051}
\bibinfo{author}{\bibfnamefont{S.~S.} \bibnamefont{Adler}} \bibnamefont{et~al.}
  (\bibinfo{collaboration}{PHENIX Collaboration}), \bibinfo{journal}{Phys. Rev.
  Lett.} \textbf{\bibinfo{volume}{96}}, \bibinfo{pages}{202301}
  (\bibinfo{year}{2006}{\natexlab{c}}).

\bibitem[{\citenamefont{Uvarov}(2001)}]{Uvarov2001}
\bibinfo{author}{\bibfnamefont{V.}~\bibnamefont{Uvarov}},
  \bibinfo{journal}{Phys. Lett.} \textbf{\bibinfo{volume}{B511}},
  \bibinfo{pages}{136} (\bibinfo{year}{2001}), \eprint{hep-ph/0105185}.

\bibitem[{\citenamefont{Cronin et~al.}(1973)}]{Cronin1}
\bibinfo{author}{\bibfnamefont{J.~W.} \bibnamefont{Cronin}}
  \bibnamefont{et~al.}, \bibinfo{journal}{Phys. Rev. Lett.}
  \textbf{\bibinfo{volume}{31}}, \bibinfo{pages}{1426} (\bibinfo{year}{1973}).

\bibitem[{\citenamefont{Cronin et~al.}(1975)}]{Cronin2}
\bibinfo{author}{\bibfnamefont{J.~W.} \bibnamefont{Cronin}}
  \bibnamefont{et~al.}, \bibinfo{journal}{Phys. Rev. D}
  \textbf{\bibinfo{volume}{11}}, \bibinfo{pages}{3105} (\bibinfo{year}{1975}).

\bibitem[{\citenamefont{Antreasyan et~al.}(1979)}]{Cronin3}
\bibinfo{author}{\bibfnamefont{D.}~\bibnamefont{Antreasyan}}
  \bibnamefont{et~al.}, \bibinfo{journal}{Phys. Rev. D}
  \textbf{\bibinfo{volume}{19}}, \bibinfo{pages}{764} (\bibinfo{year}{1979}).

\bibitem[{\citenamefont{Straub et~al.}(1992{\natexlab{a}})}]{Cronin4}
\bibinfo{author}{\bibfnamefont{P.~B.} \bibnamefont{Straub}}
  \bibnamefont{et~al.}, \bibinfo{journal}{Phys. Rev. Lett.}
  \textbf{\bibinfo{volume}{68}}, \bibinfo{pages}{452}
  (\bibinfo{year}{1992}{\natexlab{a}}).

\bibitem[{\citenamefont{Straub et~al.}(1992{\natexlab{b}})}]{Cronin5}
\bibinfo{author}{\bibfnamefont{P.~B.} \bibnamefont{Straub}}
  \bibnamefont{et~al.}, \bibinfo{journal}{Phys. Rev. D}
  \textbf{\bibinfo{volume}{45}}, \bibinfo{pages}{3030}
  (\bibinfo{year}{1992}{\natexlab{b}}).

\bibitem[{\citenamefont{Accardi et~al.}(2004)}]{Accardi2004}
\bibinfo{author}{\bibfnamefont{A.}~\bibnamefont{Accardi}} \bibnamefont{et~al.},
  \bibinfo{journal}{Phys. Lett.} \textbf{\bibinfo{volume}{B586}},
  \bibinfo{pages}{244} (\bibinfo{year}{2004}).

\bibitem[{\citenamefont{Adams et~al.}(2005{\natexlab{b}})}]{STAR_dAu1}
\bibinfo{author}{\bibfnamefont{J.}~\bibnamefont{Adams}} \bibnamefont{et~al.}
  (\bibinfo{collaboration}{STAR Collaboration}), \bibinfo{journal}{Phys. Lett.}
  \textbf{\bibinfo{volume}{B616}}, \bibinfo{pages}{8}
  (\bibinfo{year}{2005}{\natexlab{b}}).

\bibitem[{\citenamefont{Adams et~al.}(2006)}]{STAR_dAu2}
\bibinfo{author}{\bibfnamefont{J.}~\bibnamefont{Adams}} \bibnamefont{et~al.}
  (\bibinfo{collaboration}{STAR Collaboration}), \bibinfo{journal}{Phys. Lett.}
  \textbf{\bibinfo{volume}{B637}}, \bibinfo{pages}{161} (\bibinfo{year}{2006}).

\end{thebibliography}

\end{document}